\documentclass[groupedaddress,10pt,twocolumn,prl,showpacs]{revtex4}%
\usepackage{amssymb}
\usepackage{amsmath}
\usepackage{amsfonts}
\usepackage{graphicx}%
\begin{document}
\title{Violations of Locality Beyond Bell's Theorem}
\author{Zeng-Bing Chen}
\email{zbchen@ustc.edu.cn}
\affiliation{Department of Modern Physics, University of Science and Technology of China,
Hefei, Anhui 230027, China}
\author{Sixia Yu}
\affiliation{Department of Modern Physics, University of Science and Technology of China,
Hefei, Anhui 230027, China}
\author{Yong-De Zhang}
\affiliation{Department of Modern Physics, University of Science and Technology of China,
Hefei, Anhui 230027, China}

\begin{abstract}
Locality and realism are two main assumptions in deriving Bell's inequalities.
Though the experimentally demonstrated violations of Bell's inequalities rule
out local realism, it is, however, not clear what role each of the two
assumptions solely plays in the observed violations. Here we show that two
testable inequalities for the statistical predictions of two-qubit systems can
be derived by assuming either locality or realism. It turns out that quantum
mechanics respects a nonlocal classical realism, and it is locality that is
incompatible with experimental observations and quantum mechanics.

\end{abstract}
\maketitle

In their famous paper, Einstein, Podolsky and Rosen (EPR) \cite{EPR} argued
that quantum mechanics (QM) is incomplete according to their criterion of
reality based on an implicit assumption of locality. In 1964 Bell derived the
celebrated Bell inequalities (BI) \cite{Bell,CHSH,Bell-book}, enabling
quantitative tests of QM versus local realism. The derivation of BI requires
mainly the realism and locality assumptions, supplemented also by some other
auxiliary assumptions \cite{Werner-rev,assume} (e.g., no advanced actions). So
far, many experiments \cite{Aspect,Pan-GHZ} testing Bell's theorem completely
confirmed QM\ though certain technical loopholes still exist \cite{Grangier}.
Accepting the auxiliary assumptions, the \textit{experimental violations of BI
necessarily imply that at least one of the two main assumptions underlying BI
should be abandoned} \cite{Werner-rev,assume}. Then between locality and
realism, which is (or, are both) incompatible with QM?

On the one hand, the locality assumption is, at the first glance, protected by
the special theory of relativity. Its correctness can hardly be questioned. On
the other hand, realism underlies classical physics as a part of the world
view. It implies that there exists a world that is objective and
\textit{independent} of any observations. While EPR's criterion of physical
reality is certainly respected by classical physics, its status in QM seems to
be questionable due essentially to the complementarity principle. In quantum
mechanical terms, \textquotedblleft No elementary phenomenon is a phenomenon
until it is a registered (observed) phenomenon \cite{WZ-book}%
\textquotedblright, namely, what is observed on a quantum system is
\textit{dependent} upon the choice of experimental arrangements/contexts.

It should be emphasized that all experiments (the \textquotedblleft Bell
experiments\textquotedblright) performed so far to test Bell's theorem (with
\cite{Aspect} or without \cite{Pan-GHZ} inequalities) always test locality
\textit{and} realism jointly and thus, merely ruled out local realism, but
neither locality nor realism alone. Concerning the experimental violations of
BI, different attitudes arise in the literature \cite{Laloe}. Some people like
to hold the view that QM\ is nonlocal [though such a nonlocality
(\textquotedblleft\textit{Bell's nonlocality}\textquotedblright) of QM can
only be understood in the context of Bell's theorem]. For instance, Stapp
\cite{Stapp} argued that QM is a nonlocal theory. This assertion is based on
some counterfactual reasonings and arises active controversies
\cite{Stapp-comment}. Meanwhile, for other people it seems to be more natural
to give up realism \cite{inf-BI}. Thus, rejecting locality or realism becomes
again one's philosophical taste \cite{Werner-rev}, a situation very similar to
the time before the publication of Bell's work, when choosing local realism or
QM is a matter of taste \cite{Aspect}. Moreover, while biparticle entangled
pure states of any dimensionality always lead to certain violation of BI
\cite{Gisin-Peres}, the relationship between entanglement and Bell's
nonlocality for the mixed states is very puzzling \cite{Werner,hide} and
remains one of the most important open questions in the field.

In this Letter, we show that, by changing dramatically the usual way we think
about the \textquotedblleft Bell paradigm\textquotedblright, the separate role
of the locality or realism assumption \textit{can} be tested for statistical
predictions of QM by two inequalities, which are derived by only assuming
either locality or realism. For the usual Bell experiments with two orthogonal
settings per site, locality alone can lead to contradictions with experimental
observations and QM, while realistic theories can always reproduce quantum
mechanical predictions.

Obviously, to test QM versus realism and versus locality separately, a new
falsifiable formulation beyond Bell's theorem is required. To this end, one
needs first to specify the meanings of realism and locality in physical terms.
In modern understanding, EPR's criterion \cite{EPR} of realism is usually
implemented with classical hidden-variable models
\cite{Bell,Bell-book,Werner-rev,assume}. Meanwhile, locality means that the
experimental results obtained from a physical system at one location should be
independent of any observations or actions made at any other spacelike
separated locations. Previous prescription on the locality assumption was,
unfortunately, considered only within local realistic theories
\cite{Bell,Bell-book,Werner-rev,assume}. Recently, we suggested a generic
locality condition \cite{unpub} that is imposed only on\textit{\ }%
probabilities that are\textit{\ observable for localists}. The
condition\textit{\ }is\textit{\ independent upon any theory (realistic or
quantum); different theories differ only from their ways of assigning the
probabilities appearing in the locality assumption}.

The experimental configuration we have in mind is the same as that used in
deriving the usual BI. Namely, one considers an ensemble of pairs of two-level
systems A and B, which are sent, respectively, to two \textit{spacelike
separated} observers, Alice and Bob. The two-level systems can be physically
implemented by, e.g., spin-$1/2$ particles or photons with two alternative
polarizations. For definiteness, here we consider the spin-$1/2$ particles. In
QM, Alice and Bob need to measure, respectively, $\mathbf{a}\cdot
\mathbf{\sigma}^{A}\equiv\hat{a}$ and $\mathbf{b}\cdot\mathbf{\sigma}%
^{B}\equiv\hat{b}$ to obtain the statistical correlations $E(\mathbf{a}%
,\mathbf{b})$ between the two systems. Here $\mathbf{\sigma}^{A}$
($\mathbf{\sigma}^{B}$) is the Pauli spin operator of Alice's (Bob's)
particles, and $\mathbf{a}$ and$\ \mathbf{b}$ are two arbitrary unit vectors
(experimental settings). QM tells us that the observed values of $\hat{a}$,
$\hat{b}$ and $\hat{a}\hat{b}$ can only be $\pm1$ as $\hat{a}^{2}=I^{A}$,
$\hat{b}^{2}=I^{B}$ and $(\hat{a}\hat{b})^{2}=I^{A}\otimes I^{B}$, with $I$
being the unit operator. QM predicts the correlations
\begin{equation}
E_{QM}(\mathbf{a},\mathbf{b})=\mathrm{Tr}[\rho_{AB}\hat{a}\hat{b}], \label{qm}%
\end{equation}
where $\rho_{AB}$ are the states of A and B.

How does a localist interpret the observed correlations (if any)? Obviously,
events occurring in the backward light core of a particle (e.g., particle A or
B) may affect the events occurring on the particle. Particularly, events in
the overlap of the backward light cores of the two particles may be
\textquotedblleft common causes\textquotedblright\ \cite{Bell-book} of the
events occurring on A and B, though events occurring on A should not be causes
of events occurring on B (and vice versa). Denote the joint probability of
getting outcomes $a_{i}$ ($=\pm1$) and $b_{j}$ ($=\pm1$) as $P(a_{i},b_{j})$.
\textit{Whenever there are correlations in the observed }$P(a_{i},b_{j}%
)$\textit{, the localist may interpret the correlations being solely coming
from the common causes}. Then the locality assumption reads \cite{unpub}
\begin{align}
P(a_{i},b_{j})  &  =\sum_{\mu}P(a_{i},b_{j}\left\vert \mu)\right.
P(\mu)\nonumber\\
&  =\sum_{\mu}P(a_{i}\left\vert \mu)\right.  P(b_{j}\left\vert \mu)\right.
P(\mu), \label{pab}%
\end{align}
where the first line is a simple fact in theory of conditional probability.
Here the summation may also mean integration, if necessary; $P(\cdot\left\vert
\mu)\right.  $ are the probabilities conditioned on a given common
cause\ (labelled by $\mu$);\ $P(\mu)\geq0$ are the probabilities for the given
cause $\mu$ to occur, and $\sum_{\mu}P(\mu)=1$. Thus, the given common cause
can affect the probabilities with regard to particles A and B; conditioned on
the same cause, observable probabilities for A and B must be mutually
independent, as required by locality. The correlations predicted by
\textit{any local theory} (LT) are thus
\begin{equation}
E_{LT}(\mathbf{a},\mathbf{b})=\sum_{a_{i},b_{j}}a_{i}b_{j}P(a_{i},b_{j}%
)=\sum_{\mu}P(\mu)\bar{a}_{\mu}\bar{b}_{\mu}, \label{lt}%
\end{equation}
where $\bar{a}_{\mu}=\sum_{a_{i}}a_{i}P(a_{i}\left\vert \mu)\right.  $\ with
$\left\vert \bar{a}_{\mu}\right\vert \leq1$ and $\bar{b}_{\mu}=\sum_{b_{j}%
}b_{j}P(b_{j}\left\vert \mu)\right.  $ with $\left\vert \bar{b}_{\mu
}\right\vert \leq1$. If whatever the common causes are, the correlations
$E_{LT}(\mathbf{a},\mathbf{b})$ cannot be explained by local predictions
(\ref{lt}), then they are nonlocal.

The locality condition (\ref{pab}) imposes constrains merely on observable
probabilities. Particularly, localists may reasonably argue that the common
causes are not anything that is mysterious or \textquotedblleft
hidden\textquotedblright; instead, they are experimentally observable and
distinguishable (at least in principle) to account for the observed
correlations. For instance, they can be random-number generators \cite{Werner}
producing numbers $\mu$ with probabilities $P(\mu)$, which are held in a
preparing device creating the statistical ensemble under study. In this way,
one can exclude any assumption other than locality. A theory has the power of
making predictions. Thus, using either QM or RT, one can predict the
probabilities in (\ref{pab}). There are then two other facts supporting
(\ref{pab}) as a generic locality condition.

First, we proved \cite{unpub} recently that for spacelike separated systems,
Eq. (\ref{pab}) is obeyed iff the states of the two particles are separable
[i.e., entangled states possess quantum nonlocality in the sense of violating
the locality condition (\ref{pab})]. Thus, a local quantum theory (LQT, i.e.,
QM$+$locality) predicts
\begin{equation}
E_{LQT}(\mathbf{a},\mathbf{b})=\sum_{\mu}P(\mu)\mathrm{Tr}\left[  \rho_{A\mu
}\hat{a}\right]  \mathrm{Tr}[\rho_{B\mu}\hat{b}], \label{lqt}%
\end{equation}
where $\rho_{A\mu}$ ($\rho_{B\mu}$) are the local density operators
conditioned on the common cause $\mu$ such that $\bar{a}_{\mu}=\mathrm{Tr}%
\left[  \rho_{A\mu}\hat{a}\right]  $ and $\bar{b}_{\mu}=\mathrm{Tr}[\rho
_{B\mu}\hat{b}]$ [see Eq. (\ref{lt})].

Second, if the two particles in question are described by a classical
realistic theory (RT), Eq. (\ref{pab}) becomes Bell's locality condition which
has been well justified in various aspects in the context of BI
\cite{Bell-book,Werner-rev,assume} and now is widely accepted. To see this,
recall that in an RT, the probabilities in (\ref{pab}) are determined by the
experimental settings and by a set of hidden variables, denoted collectively
by $\lambda$, and as such $P(a_{i},b_{j})=\sum_{\mu}\int d\lambda
p(\lambda)P_{\lambda}(a_{i}\left\vert \mu)\right.  P_{\lambda}(b_{j}\left\vert
\mu)\right.  P_{\lambda}(\mu)$, where $p(\lambda)\geq0$ is a normalized
probability distribution of $\lambda$.\ If one \textit{formally} identifies
the common causes as a part of the hidden variables, then Bell's locality
condition \cite{Bell-book,Werner-rev,assume}\ can be obtained. The
correlations predicted by local realistic theories (LRT) are then
\begin{equation}
E_{LRT}(\mathbf{a},\mathbf{b})=\sum_{\mu}\int d\lambda p(\lambda)P_{\lambda
}(\mu)A_{\mu}(\mathbf{a},\lambda)B_{\mu}(\mathbf{b},\lambda), \label{lrt}%
\end{equation}
where $A_{\mu}(\mathbf{a},\lambda)=\sum_{a_{i}}a_{i}P_{\lambda}(a_{i}%
\left\vert \mu)\right.  $\ with $\left\vert A_{\mu}(\mathbf{a},\lambda
)\right\vert \leq1$ and $B_{\mu}(\mathbf{b},\lambda)=\sum_{b_{j}}%
b_{j}P_{\lambda}(b_{j}\left\vert \mu)\right.  $ with $\left\vert B_{\mu
}(\mathbf{b},\lambda)\right\vert \leq1$. However, an RT without assuming
locality predicts
\begin{equation}
E_{RT}(\mathbf{a},\mathbf{b})=\sum_{\mu}\int d\lambda p(\lambda)P_{\lambda
}(\mu)\Gamma_{\mu}(\mathbf{a},\mathbf{b};\lambda), \label{rt}%
\end{equation}
with $\Gamma_{\mu}(\mathbf{a},\mathbf{b};\lambda)=\sum_{a_{i},b_{j}}a_{i}%
b_{j}P_{\lambda}(a_{i},b_{j}\left\vert \mu)\right.  $ and $\left\vert
\Gamma_{\mu}(\mathbf{a},\mathbf{b};\lambda)\right\vert \leq1$.

Clearly, the \textit{same} locality assumption (\ref{pab}) is underlying both
(\ref{lrt}) and (\ref{lqt}), where QM and realism differ from their distinct
ways of assigning probabilities (or measured results) for the same quantities.
Now our task is to deduce the consequences for each of the five theories (LT,
RT, LRT, LQT and QM) in the Bell experiments, to see if there are testable
quantitative differences among them.

To obtain the required inequalities, Alice (Bob) needs to measure at another
direction $\mathbf{a}_{\perp}\perp\mathbf{a}$ ($\mathbf{b}_{\perp}%
\perp\mathbf{b}$), namely, we are concerned with the inequalities with two
\textit{orthogonal} settings per site. Then consider the following
combinations of the correlation functions:
\[
E(\mathbf{a},\mathbf{b}_{\perp})+E(\mathbf{a}_{\perp},\mathbf{b})\equiv
X,\ \ E(\mathbf{a},\mathbf{b})-E(\mathbf{a}_{\perp},\mathbf{b}_{\perp})\equiv
Y.
\]
In terms of $X$ and $Y$ QM\ predicts the following inequality for all two-spin
states (entangled or not) \cite{Uffink-Yu}
\begin{equation}
X_{QM}^{2}+Y_{QM}^{2}\leq4, \label{yu}%
\end{equation}
which can, actually, be proved by using the Heisenberg-Robertson uncertainty
relation for the composite system. Using the facts that $\left\vert
\Gamma_{\mu}(\mathbf{a},\mathbf{b};\lambda)\right\vert $, $\left\vert
\Gamma_{\mu}(\mathbf{a}_{\perp},\mathbf{b}_{\perp};\lambda)\right\vert $,
$\left\vert \Gamma_{\mu}(\mathbf{a}_{\perp},\mathbf{b};\lambda)\right\vert $,
$\left\vert \Gamma_{\mu}(\mathbf{a},\mathbf{b}_{\perp};\lambda)\right\vert
\leq1$, it can be proved that $\left\vert X_{RT}\right\vert \leq\sum_{\mu}\int
d\lambda p(\lambda)P_{\lambda}(\mu)\left[  \left\vert \Gamma_{\mu}%
(\mathbf{a},\mathbf{b}_{\perp};\lambda)\right\vert +\left\vert \Gamma_{\mu
}(\mathbf{a}_{\perp},\mathbf{b};\lambda)\right\vert \right]  \leq2$, and
similarly $\left\vert Y_{RT}\right\vert \leq2$. Thus, the inequality imposed
by realism alone is
\begin{equation}
\left\vert X_{RT}\right\vert \leq2,\ \ \ \ \left\vert Y_{RT}\right\vert \leq2.
\label{RT2}%
\end{equation}
Note that in a RT, all the four $\Gamma_{\mu}$ functions used above can be
mutually independent. So no lower bound exists for Eq. (\ref{RT2}).

From (\ref{lt}), one can obtain the \textquotedblleft locality
inequality\textquotedblright\ satisfied by any local theory
\begin{equation}
\left\vert X_{LT}\pm Y_{LT}\right\vert \leq2 \label{local}%
\end{equation}
due to the fact that $\left\vert (\bar{a}_{\mu}\pm\bar{a}_{\perp\mu})\bar
{b}_{\mu}+(\bar{a}_{\mu}\mp\bar{a}_{\perp\mu})\bar{b}_{\perp\mu}\right\vert
\leq2$. Particularly, the BI (due to Clauser, Horne, Shimony and Holt
\cite{CHSH}) imposed by any local realistic theory reads
\begin{equation}
\left\vert X_{LRT}\pm Y_{LRT}\right\vert \leq2, \label{chsh}%
\end{equation}
for which no tighter bound exists.

However, an LQT\ predicts an inequality (the \textquotedblleft quantum
locality inequality\textquotedblright)
\begin{equation}
X_{LQT}^{2}+Y_{LQT}^{2}\leq1,\label{LQT1}%
\end{equation}
which is \textit{stronger} than the locality inequality (\ref{local}). The
proof of the inequality (\ref{LQT1}) is easy. Using Eq. (\ref{lqt}) and the
property \cite{Uffink-Yu} of $X_{LQT}^{2}$ and $Y_{LQT}^{2}$ being convex
functions of local density operators, it suffices to prove the validity of
(\ref{LQT1}) for $\rho_{AB}=\rho_{A}\rho_{B}$. Denoting $\mathrm{Tr}\left[
\rho_{A}\hat{a}\right]  =\left\langle \hat{a}\right\rangle _{A}$ and
$\mathrm{Tr}[\rho_{B}\hat{b}]=\langle\hat{b}\rangle_{B}$, one has $X_{LQT}%
^{2}+Y_{LQT}^{2}=(\left\langle \hat{a}\right\rangle _{A}^{2}+\left\langle
\hat{a}_{\perp}\right\rangle _{A}^{2})(\langle\hat{b}\rangle_{B}^{2}%
+\langle\hat{b}_{\perp}\rangle_{B}^{2})\leq1$, where $\left\langle \hat
{a}\right\rangle _{A}^{2}+\left\langle \hat{a}_{\perp}\right\rangle _{A}%
^{2}\leq1$ and $\langle\hat{b}\rangle_{B}^{2}+\langle\hat{b}_{\perp}%
\rangle_{B}^{2}\leq1$ have been exploited and are direct consequences of the
Heisenberg-Robertson uncertainty relation for each subsystem A/B. In the
present case the uncertainty relation gives, e.g., for Alice's particle
$(1-\left\langle \hat{a}\right\rangle _{A}^{2})(1-\left\langle \hat{a}_{\perp
}\right\rangle _{A}^{2})\geq\left\langle (\mathbf{a\times a}_{\perp}%
)\cdot\mathbf{\sigma}_{A}\right\rangle _{A}^{2}+\left\langle \hat
{a}\right\rangle _{A}^{2}\left\langle \hat{a}_{\perp}\right\rangle _{A}^{2}$,
yielding $\left\langle \hat{a}\right\rangle _{A}^{2}+\left\langle \hat
{a}_{\perp}\right\rangle _{A}^{2}\leq\left\langle \hat{a}\right\rangle
_{A}^{2}+\left\langle \hat{a}_{\perp}\right\rangle _{A}^{2}+\left\langle
(\mathbf{a\times a}_{\perp})\cdot\mathbf{\sigma}_{A}\right\rangle _{A}^{2}%
\leq1$.%
%TCIMACRO{\FRAME{ftbpFU}{2.2096in}{2.222in}{0pt}{\Qcb{Five inequalities
%(\ref{yu}-\ref{LQT1}) in the $X$-$Y$ plane. They are predicted by: LT/LRT (the
%inner tilted square), QM (the outer circle with radius $2$), RT (the outer
%horizontal square) and LQT (the inner circle with radius $1$).}}{\Qlb{fig}%
%}{bfig.eps}{\special{ language "Scientific Word";  type "GRAPHIC";
%width 2.2096in;  height 2.222in;  depth 0pt;  original-width 2.9089in;
%original-height 2.9238in;  cropleft "0";  croptop "1";  cropright "1";
%cropbottom "0";  filename 'bfig.eps';file-properties "XNPEU";}}}%
%BeginExpansion
\begin{figure}
[ptb]
\begin{center}
\includegraphics[
height=2.222in,
width=2.2096in
]%
{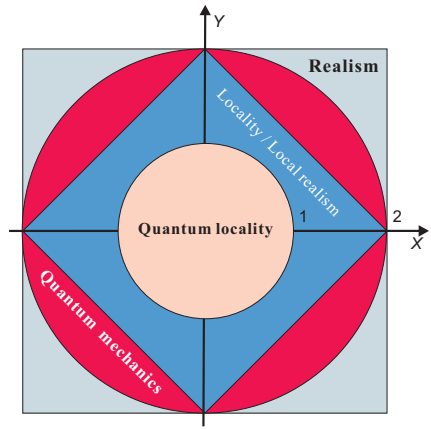}%
\caption{Five inequalities (\ref{yu}-\ref{LQT1}) in the $X$-$Y$ plane. They
are predicted by: LT/LRT (the inner tilted square), QM (the outer circle with
radius $2$), RT (the outer horizontal square) and LQT (the inner circle with
radius $1$).}%
\label{fig}%
\end{center}
\end{figure}
%EndExpansion

Since the inequalities (\ref{yu}-\ref{LQT1}) make their predictions on the
same experiments, they can be summarized in a single diagram shown in Fig. $1
$. Seen from the diagram, there is an interesting relation among the
predictions of the five theories
\begin{equation}
\text{RT}\supset\text{QM}\supset\text{LT}/\text{LRT}\supset\text{LQT.}
\label{hie}%
\end{equation}
Note that the inequalities (\ref{yu}-\ref{chsh}) and, thus, the relation
RT$\supset$QM$\supset$LT/LRT are still valid for the cases where the two
settings for each site are not orthogonal. The relation (\ref{hie}) implies
that, e.g., all QM predictions are also predicted by RT though being
essentially classical, but the RT predicts something more than QM. Thus, for
the system in question all predictions do allow to be interpreted by certain
classical hidden-variable model, which must be of a nonlocal nature as a price
\cite{Werner-rev}. In other words, realism (without assuming locality) in
itself is not excluded by QM. However, accepting that QM\ is correct (and very
unlikely to be wrong for systems as simple as two-level ones), it is
impossible to observe the conflict between RT and QM in the Bell experiments.

The relation (\ref{hie}) definitely shows that there are \textit{two
qualitatively different nonlocality}. Any observed correlation that does not
satisfy the locality condition (\ref{pab}) is nonlocal. When the probabilities
in (\ref{pab}) are quantum predictions, (\ref{pab}) can be reasonably called
the \textquotedblleft quantum locality condition\textquotedblright, whose
violation indicates quantum nonlocality (\textquotedblleft
qunonlocality\textquotedblright\ for short). It is qunonlocality that is
proved to be equivalent to entanglement for spacelike separate quantum systems
\cite{unpub}. The fact that the locality inequality (\ref{local}) and the BI
(\ref{chsh}) take the same form implies that the Bell experiments performed to
test (\ref{chsh}) actually ruled out all local theories (including LRT and
LQT) and proved nonlocality of nature for statistical predictions of QM. The
distinct trends of locality and realism in Fig. $1$ show that realism can mask
qunonlocality. We think this is the reason why the relation between
nonlocality and entanglement is such a notoriously difficult issue when being
seen in the context of BI. Indeed, there exist two-particle entangled states
(the Werner states \cite{Werner}) with \textquotedblleft hidden
nonlocality\textquotedblright\ \cite{hide} (more precisely, hidden
qunonlocality), hidden by realism so much that it cannot be uncovered by any
BI. All states in the region constrained by $\left\vert X_{LRT}\right\vert
+\left\vert Y_{LRT}\right\vert \leq2$ and $X_{LQT}^{2}+Y_{LQT}^{2}>1$\ in Fig.
$1$ possess hidden qunonlocality.

Let us consider the experimental settings under which the locality inequality
(\ref{LQT1}) is violated by QM. It is well known that the bound of BI
(\ref{chsh}) [and thus, (\ref{local})] allowed by QM is $2\sqrt{2}$, known as
the Cirel'son bound \cite{Cirelson}. However, the $2$-setting quantum locality
inequality (\ref{LQT1}) has the bound $4$ of maximal violation [see
(\ref{yu})] and can reveal a much stronger violation allowed by QM.

Choose the spin singlet state $\left\vert \psi^{-}\right\rangle =\frac
{1}{\sqrt{2}}(\left\vert \uparrow\right\rangle _{A}\left\vert \downarrow
\right\rangle _{B}-\left\vert \downarrow\right\rangle _{A}\left\vert
\uparrow\right\rangle _{B})$, where $\left\vert \uparrow\right\rangle $
($\left\vert \downarrow\right\rangle $) is the spin-up
(spin-down)\ state.\ Then $E_{QM}(\mathbf{a},\mathbf{b})=\left\langle \psi
^{-}\right\vert \hat{a}\hat{b}\left\vert \psi^{-}\right\rangle =-\mathbf{a}%
\cdot\mathbf{b}$ gives $X_{QM}^{2}+Y_{QM}^{2}=(\mathbf{a}\cdot\mathbf{b}%
_{\perp}+\mathbf{a}_{\perp}\cdot\mathbf{b})^{2}+(\mathbf{a}\cdot
\mathbf{b}-\mathbf{a}_{\perp}\cdot\mathbf{b}_{\perp})^{2}$. The maximal bound
$4$ allowed by QM\ can easily be attained by choosing the angles from
$\mathbf{a}_{\perp}$\ to $\mathbf{b}_{\perp}$\ and from $\mathbf{b}_{\perp}%
$\ to $\mathbf{a}_{\perp} $\ to be $\pi/4$. Then for the Werner state
\cite{Werner,Peres,Horodecki} $\rho_{W}=\frac{1}{4}(1-x)+x\left\vert \psi
^{-}\right\rangle \left\langle \psi^{-}\right\vert \ $(here $1>x>0$) and the
same settings chosen above, $\left\vert X_{QM}\right\vert +\left\vert
Y_{QM}\right\vert \leq2\sqrt{2}x$ implies that when $x>1/\sqrt{2}$,
(\ref{local}) or (\ref{chsh}) is violated by QM.\ It is already known that for
$x>1/3$ the Werner state $\rho_{W}$\ is entangled \cite{Peres,Horodecki},
i.e., has qunonlocality. Therefore, there is hidden qunonlocality for
$1/\sqrt{2}\geq x>1/3$ that does not lead to any violation of BI (\ref{chsh})
or the locality inequality (\ref{local}). Meanwhile, under the same condition
$X_{QM}^{2}+Y_{QM}^{2}=4x^{2}$; when $x>1/2$, the quantum locality inequality
(\ref{LQT1}) is violated by $\rho_{W}$. Thus, (\ref{LQT1}) shows a sharper
contradiction with QM than (\ref{local}) or (\ref{chsh}).

The fact that the quantum locality inequality (\ref{LQT1}) cannot fully
uncover qunonlocality in $\rho_{W}$ means that its violation is only a
sufficient condition for qunonlocality, but not a necessary one. Yet, the
quantum counterpart of the locality condition (\ref{pab}) is necessary and
sufficient for separability of states for spacelike separated systems and
thus, all entangled states possess qunonlocality \cite{unpub}. We mention that
a necessary and sufficient condition of separability of states for the Bell
experiments with three mutually orthogonal settings per site was found
recently for two-qubit systems \cite{Yu-detect}.

To summarize, we have established a hierarchy [see Eq. (\ref{hie}) and Fig.
$1$] of five kinds of theories (RT, QM, LT, LRT and LQT) for the usual Bell
experiments with two orthogonal settings per site. The hierarchy enables
separate experimental tests of QM versus locality beyond Bell's theorem. It
also sheds new light on the role of locality or realism in the experimental
violations of BI and the relationship between entanglement and Bell's
nonlocality. The quantum locality inequality is useful for detecting genuine
qunonlocality and might find interesting applications in quantum information
processing. For instance, for the EPR protocol of quantum cryptography
\cite{Ekert}, (\ref{LQT1}) may lead to better test of eavesdropping.
Interestingly, violation of locality without inequalities for multiparticle
Greenberger-Horne-Zeilinger states \cite{GHZ-90} can also be proved and will
be reported elsewhere. Finally, we stress that (quantum) nonlocality (or,
entanglement) cannot be used for superluminal communication \cite{Laloe,unpub}.

We thank Jian-Wei Pan and Nai-Le Liu for stimulating discussions. This work
was supported by the National NSF of China under Grant No. 10104014, the CAS
and the National Fundamental Research Program under Grant No. 2001CB309300.


\begin{thebibliography}{99}                                                                                               %


\bibitem {EPR}A. Einstein, B. Podolsky, and N. Rosen, Phys. Rev. \textbf{47},
777 (1935).

\bibitem {Bell}J.S. Bell, Physics (Long Island City, N.Y.) \textbf{1}, 195 (1964).

\bibitem {CHSH}J.F. Clauser, M.A. Horne, A. Shimony and R.A. Holt, Phys. Rev.
Lett. \textbf{23,} 880 (1969).

\bibitem {Bell-book}J.S. Bell, \textit{Speakable and Unspeakable in Quantum
Mechanics} (Cambridge University Press, Cambridge, 1987).

\bibitem {Werner-rev}R.F. Werner and M.M. Wolf, Quantum Inf. Comput.
\textbf{1} (3), 1 (2001).

\bibitem {assume}A.G. Valdenebro, Eur. J. Phys. \textbf{23}, 569 (2002).

\bibitem {Aspect}A. Aspect, Nature (London) \textbf{398}, 189 (1999).

\bibitem {Pan-GHZ}J.-W. Pan \textit{et al}., Nature (London) \textbf{403}, 515 (2000).

\bibitem {Grangier}P. Grangier, Nature (London) \textbf{409}, 774 (2001).

\bibitem {WZ-book}J.A. Wheeler and W.H. Zurek, \textit{Quantum Theory and
Measurement} (Princeton University Press, Princeton, New Jersey, 1983).

\bibitem {Laloe}F. Lalo\"{e}, Am. J. Phys. \textbf{69}, 655 (2001).

\bibitem {Stapp}H.P. Stapp, Am. J. Phys. \textbf{65}, 300 (1997).

\bibitem {Stapp-comment}See, e.g., N.D. Mermin, Am. J. Phys. \textbf{66}, 920
(1998); A. Shimony and H. Stein, \textit{ibid}. \textbf{69}, 848 (2001); H.P.
Stapp, \textit{ibid}. \textbf{69}, 854 (2001); W. Unruh, Phys. Rev. A
\textbf{59}, 126 (1999); H.P. Stapp, \textit{ibid}. A \textbf{60}, 2595
(1999); W. Unruh, \textit{ibid}. A \textbf{60}, 2599 (1999).

\bibitem {inf-BI}S.L. Braunstein and C.M. Caves, Phys. Rev. Lett. \textbf{61},
662 (1988).

\bibitem {Gisin-Peres}N. Gisin and A. Peres, Phys. Lett. A \textbf{162}, 15 (1992).

\bibitem {Werner}R.F. Werner,\ Phys. Rev. A \textbf{40}, 4277 (1989).

\bibitem {hide}S. Popescu, Phys. Rev. Lett. \textbf{74}, 2619 (1995); N.
Gisin, Phys. Lett. A \textbf{210}, 151 (1996); M. \.{Z}ukowski \textit{et
al}., Phys. Rev. A \textbf{58}, 1694 (1998); A. Peres, \textit{ibid}.
\textbf{54}, 2685 (1996).

\bibitem {unpub}Z.-B. Chen, S. Yu, Y.-D. Zhang, and N.-L. Liu, to be published.

\bibitem {Uffink-Yu}J. Uffink,\ Phys. Rev. Lett. \textbf{88}, 230406 (2002);
S. Yu, Z.-B. Chen, J.-W. Pan, and Y.-D. Zhang, \textit{ibid}. \textbf{90}, 080401.

\bibitem {Cirelson}B.S. Cirel'son, Lett. Math. Phys. \textbf{4}, 93 (1980).

\bibitem {Peres}A. Peres, Phys. Rev. Lett. \textbf{77}, 1413 (1996).

\bibitem {Horodecki}M. Horodecki \textit{et al}., Phys. Lett. A\textbf{\ 223,}
1 (1996).

\bibitem {Yu-detect}S. Yu, J.-W. Pan, Z.-B. Chen, and Y.-D. Zhang, quant-ph/0301030.

\bibitem {Ekert}A.K. Ekert, Phys. Rev. Lett. \textbf{67}, 661 (1991).

\bibitem {GHZ-90}D.M. Greenberger, M.A. Horne, A. Shimony, and A. Zeilinger,
Am. J. Phys. \textbf{58}, 1131 (1990).
\end{thebibliography}
\end{document}